\documentclass[12pt]{iopart}

\usepackage{graphicx}

\begin{document}

\title[Sensitivity, entropy, and escape rates at the onset of chaos]
{Sensitivity to initial conditions, entropy production, \\
and escape rate at the onset of chaos}

\author{Miguel Angel Fuentes$^{1,2,3}$, Yuzuru Sato$^{4}$ and Constantino Tsallis$^{1,5}$}

\address{$^1$Santa Fe Institute, 1399 Hyde Park Road, Santa Fe, New Mexico 87501, USA}
\address{$^2$Centro At\'omico Bariloche, Instituto Balseiro and CONICET, 8400 Bariloche, Argentina}
\address{$^3$ Center for Advanced Studies in Ecology and Biodiversity, Facultad de Ciencias Biol\'ogicas, Pontificia Universidad Cat\'olica de Chile, Casilla 114-D, Santiago CP 6513677, Chile}
\address{$^4$ RIES, Department of Mathematics, Hokkaido University, Kita 20 Nishi 10, Kita-ku, Sapporo, Hokkaido 001-0020, Japan}
\address{$^5$Centro Brasileiro de Pesquisas Fisicas and National Institute of Science and Technology for Complex Systems, Rua Xavier Sigaud 150, 22290-180 Rio de Janeiro-RJ, Brazil}

\ead{fuentesm@santafe.edu}
\begin{abstract}
We analytically link three properties of nonlinear dynamical
systems, namely sensitivity to initial conditions, entropy production, and escape rate,
in $z$-logistic maps for both positive and zero Lyapunov exponents. We unify
these relations at chaos, where the Lyapunov exponent is positive, and at its onset,
where it vanishes. Our result unifies, in particular, two already known cases, namely (i)
the standard entropy rate in the presence of escape, valid for exponential functionality
rates with strong chaos, and (ii) the Pesin-like
identity with no escape, valid for the power-law behavior present at
points such as the Feigenbaum one.

\end{abstract}

\maketitle

\section{Introduction}

At the onset of chaos, some traditional approaches to understand specific dynamical behavior, such as the entropy production, do not provide quantitatively nontrivial information on the state of the system. From the various physical phenomena which, at this limit, show a clear departure from the classical Boltzmann-Gibbs (BG) theory, the edge of chaos of one-dimensional maps has played the role of an archetypical system of study (see, for example, \cite{libros,GellMannTsallis2004,Tsallis2009} and references therein). In this manuscript we will focus on the relation between sensitivity to initial conditions, a possible escape within traps, and the loss of information in an one-dimensional map at the onset of chaos.

\subsection{The $z$-logistic map}

Let us focus on logistic-like maps $x_{t+1}=f(x_t)$ ($t=0,1,2,...$) with
\begin{equation}
f(x) =  1-\mu |x|^z\;\;(z>1;~ 0 \le \mu \le 2;~ -1 \le x \le 1). \label{map}
\end{equation}

The expression of the sensitivity to initial conditions $\xi_t \equiv \lim_{\Delta x_0 \to 0} \frac{\Delta x_t}{\Delta x_0}$ has
been derived using exact analytic renormalization group theory (RGT), leading to
\begin{eqnarray}
\xi_t & = & e_{q_{sen}}^{\lambda_{q_{sen}}\,t} = [1+(1-q_{sen}) \lambda_{q_{sen}}t]^{\frac{1}{1-q_{sen}}}, \label{one}
\end{eqnarray}
with $q_{sen} \le 1$, and $e_q^y \equiv [1+(1-q)y]^{\frac{1}{1-q}}$ ($e_1^y = e^y$). When the Lyapunov exponent $\lambda_1$ is positive, we have $q_{sen}=1$. At the edge of chaos, from \cite{BaldovinRobledo2002}, we have
\begin{equation}
\lambda_{q_{sen}}=\frac{1}{1-{q_{sen}}}=\frac{(z-1)\ln \alpha(z)}{\ln 2} \,,
\label{lambda}
\end{equation}
where $\alpha(z)$ is the $z$-generalized Feigenbaum universal constant (see also \cite{AnaniaPoliti1988}). In particular
$\alpha(2)=2.50290...$.
It follows that $\xi_t=(1+t)^{\lambda_{q_{sen}}}\sim t^{\lambda_{q_{sen}}}\;\;(t\to\infty)$.

In this context, the nonadditive entropy  \cite{Tsallis1988}
\begin{eqnarray}
S_q  &=&  \sum_i p_i \ln_q \left( \frac{1}{p_i} \right) \equiv \frac{1-\sum_i p_i^q}{q-1}  \label{qentro} \\
(S_1 &=&S_{BG} \equiv  -\sum_ip_i \ln p_i), \nonumber
\end{eqnarray}
has been successfully used \cite{LyraTsallis1998}, where $BG$ stands for {\it Boltzmann-Gibbs}, and $\ln_q y \equiv \frac{y^{1-q}-1}{1-q}$ ($\ln_1 y=\ln y$).

\section{Pesin's theorem}

More recently, using RGT, it has been proved an extension of Pesin's theorem \cite{BaldovinRobledo2004}, valid for the entire family of these unimodal maps, making the important connection between the loss of information measure, via entropy production
\begin{eqnarray}
K_{q_{ent}} &\equiv& \lim_{t \to\infty} \frac{S_{q_{ent}}}{t}\,,
\label{definition}
\end{eqnarray}
and
the sensitivity to initial conditions. This result can be expressed by the $q$-generalized Pesin-like identity
\begin{eqnarray}
K_{q_{ent}} & = & \lambda_{q_{sen}},
\label{pesinequality}
\end{eqnarray}
with $q_{ent}=q_{sen}$. Whenever $\lambda_1 >0$  we have
\begin{equation}
q_{ent}=q_{sen}=1.
\end{equation}
At the Feigenbaum point $\mu_\infty$ (where $\lambda_1 =0$), Eq. (\ref{lambda}) is satisfied.
For $z=2$ we have $\mu_\infty= 1.401155...$ (for $z$ increasing from 1 to infinity, $\mu_\infty(z)$ monotonically increases from 1 to 2). At this point, the pitchfork bifurcations accumulate, and the transition to chaos occurs.

It is worth remembering that the Pesin equality relates the Lyapunov exponent $\lambda_1$
of nonlinear maps to the Kolmogorov-Sinai (KS) entropy $\mathcal{K}$.
For relation (\ref{pesinequality}),  the probability distribution in Eq. (\ref{qentro}) is
obtained as the frequency at which the $i$-th cell in the state space,
$i=1,2,...$, is occupied at a given time.
The main difference that one would expect between a possible $q$-generalization of the Kolmogorov-Sinai rate, noted $\mathcal{K}_q$, and the entropy production
rate $K_q$, in Eq. (\ref{pesinequality}), is that in the latter the initial conditions for
the trajectories have a specific distribution $p_i(0)$ while the $q$-KS-entropy case, $\mathcal{K}_q$,
considers all the trajectories from their initial positions to the time
$t\rightarrow\infty$. The connection between $K_1$ and
$\mathcal{K}$ has been studied for several chaotic maps \cite{LatoraBaranger1999} and
it has been determined that typically the asymptotic equality $K_1=\mathcal{K}$ holds.
In numerical calculations, this corresponds to an intermediate stage of evolution, after a transient and before
reaching the asymptotic (saturated) state. In the same manner, we are interested here in the temporal
region where $\mathcal{K}_q=K_q$, and, as shown in \cite{BaldovinRobledo2004},
it will be necessary to study only uniform initial distribution.

We introduce the definition of the sensitivity factor $\xi_{t_k}$ for each
time subsequence $k$, $k=1,2,3...$, and for two trajectories $x_{t_k}$ and
$y_{t_k}$ with initial condition $x_{in}$, $y_{in}$
\begin{eqnarray}
\xi_{t_k} & = & \lim_{|y_{in}-x_{in}| \rightarrow 0} \frac{\left|
y_{t_k}(y_{in})- x_{t_k}(x_{in}) \right|}{\left| y_{t_k =0}(y_{in})- x_{t_k
=0}(x_{in})  \right|}. \nonumber
\end{eqnarray}

We will consider in this map a partition of $W$ equal-length boxes of size
$\Delta$, and an ensemble of $N$ trajectories with initial condition near
$x_{in}=1$ (in what follows, we will omit the $k$-index, that, as we mentioned,
describes a possible time subsequence). Then, the entropy production can be calculated considering the occupancy frequency of each box.
The total number of cells that the ensembles of trajectories
occupy at time $t$ will be $W_t=\Delta_t/\Delta$, where $\Delta_t$ is the
total size occupied by the ensemble. As pointed out in \cite{BaldovinRobledo2004}, in
the limit of $\Delta\rightarrow0$, the total number of occupied cells will be
uniformly distributed as $W_t=\xi_t$. This result implies that the
total entropy at time $t$ will be simply given by $S_q=\ln_q W_t=\lambda_q~t$,
leading to Eq. (6).
In our analysis we will consider the same type of expansion for a
generic dynamic map, that is
\begin{eqnarray}
W_{t}=\xi_{t}. \label{w_t}
\end{eqnarray}
This relation is based on maps that have been studied in the literature, as
discussed above, some of them homomorphism of important continuous chaotic
dynamical systems \cite{Ott2002}. The key point in our discussion will be the
relation between this expansion and the functionality of the escape rate.

\section{Open systems}

In the presence of escape gates in the state space, the exponential decay behavior (corresponding to $\lambda_1>0$) for a given
number of initial trajectories in the phase space has been
extensively reported \cite{libros}. This means that for a given number of initial
trajectories, the number of them that remain in the system at time $t$, $N_t$,
will be given by
\begin{eqnarray}
N_t=N_0\, e^{-\gamma_1 ~ t}.
\end{eqnarray}
For a dynamical system that has an exponential uniform expansion, and under the
presence of an escape rate that behaves as the previously mentioned one, the
number of occupied cells at time $t$ can be expressed as
\begin{eqnarray}
W_{t}=e^{-\gamma_1 ~ t} e^{\lambda_1 ~ t}.
\end{eqnarray}
And using Eq. (\ref{qentro}), with $q = 1$, we can obtain the well known relation
\begin{eqnarray}
S_1 =\ln [W_t]=(\lambda_1-\gamma_1)t \,.
\end{eqnarray}
Consequently, given that $K_1 \equiv \lim_{t \to\infty}\frac{S_1}{t}$, we have
\begin{eqnarray}
K_1 =\lambda_1-\gamma_1,
\label{relation}
\end{eqnarray}
which connects, for such systems, the entropy production $K_1$, the
Lyapunov exponent $\lambda_1$, and the escape rate $\gamma_1$.

Notice that in principle the functionality of both the divergence of
initial conditions and the escape should be the same in
order for this relation to exist (exponential in the strong chaos, power-law at its onset).

Consider now an uniform expansion that behaves as a power law (instead of as an
exponential), for example, the logistic map at $\mu_\infty$: what will the
entropy production be under a power law-type of escape? Note that, if the
decreasing number of trajectories in the presence of escape is exponential,
the behavior of the occupied bins after some time will be far from power law.

Assuming that our system is such that the escape asymptotically follows
\begin{eqnarray}
N_{t}\propto t^{-\gamma_{q_{esc}}}\,,
\label{exponentescape}
\end{eqnarray}
and taking into account that
\begin{equation}
N_t=N_0\, e_{q_{esc}}^{-\gamma_{q_{esc}}\,t} \propto t^{-1/(q_{esc}-1)}  \;\;\;(t \to\infty)\,,
\end{equation}
we can identify
\begin{equation}
\gamma_{q_{esc}}= \frac{1}{q_{esc}-1} \,.
\end{equation}

The production of occupied bins, $W_{t}$, and sensibility $\xi_{t}$ (see Eq. (\ref{w_t}))   will behave as
\begin{eqnarray}
W_{t}=\xi_{t}\propto t^{\lambda_{q_{sen}}} t^{-\gamma_{q_{esc}}}. \label{w_of_t}
\end{eqnarray}
Then, finally, the entropy production (from Eq. (\ref{definition})) is given by
\begin{equation}
K_{q_{ent}}  = \lambda_{q_{sen}} -\gamma_{q_{esc}} \, ,
\label{pesinEscape}
\end{equation}
with
\begin{equation}
K_{q_{ent}}= \frac{1}{1-q_{ent}}\,.
\label{q_pesinEscape2}
\end{equation}
Eq. (\ref{pesinEscape}) generalizes Eq. (\ref{relation}), the well known relation valid for chaotic systems (i.e., $\lambda_1>0$) in the presence of escape (see for instance \cite{libros}). Notice that the $q$ indices present in Eq. (\ref{pesinEscape}) do not need to be equal among them. Indeed the exponents in the power laws shown in Eqs. (\ref{one}) and (\ref{exponentescape}) do not necessarily coincide. This implies, through Eq. (\ref{q_pesinEscape2}), that also $q_{ent}$ generically differs from them.

\subsection{Example}

As a concrete example of a system that has the properties discussed above, we can mention the following non-autonomous map
\begin{eqnarray}
x_{t+1} &=& [1-\mu (x_t^2+y_t^2)] \cos\Theta_t \nonumber \\
y_{t+1} &=& [1-\mu (x_t^2+y_t^2)] \sin\Theta_t,
\end{eqnarray}
with
\begin{eqnarray}
\Theta_t=\pi \frac{\ln~(t+1)}{\ln~2},
\end{eqnarray}
which is effectively a rotating logistic map. The values of $\cos\Theta_i$ ($\sin\Theta_i$) will be $\pm 1$ (0) exactly when the map expands in the $x$ axis (with $y=0$). This system can simultaneously show a critical behavior such as that at the onset of chaos (i.e., at $\mu_{\infty}$), and power-law escaping due to holes that are uniformly distributed on the line $y=0$. In such situation we can achieve an exact solution for the entropy production, using the usual ensemble of initial trajectories at $y=0$, near $x=1$ \cite{BaldovinRobledo2004}, with $t=2^i-1$, $i=1,2,3...$.

\begin{figure}[h]
\centering \includegraphics[width=8.5cm,angle=0]{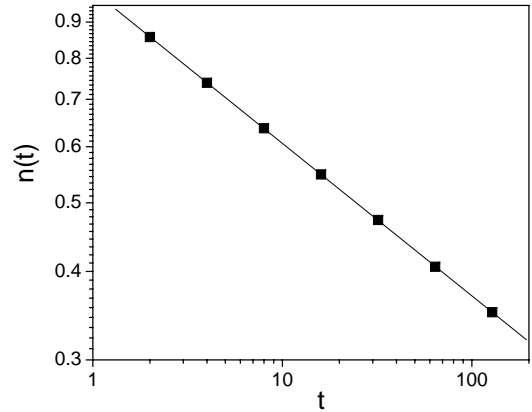}\\
\caption{Fraction of points, $n(t)=N_0/N_t$, remaining in the system versus time using $\delta=6/7 \sim 0.86$ and $z=2$, in log-log scale. $N_0=10^6$ uniformly taken within the interval $[1-10^{-10}, 1]$.  The fit, dashed line, shows a escape parameter $\gamma_{q_{esc}} = 0.216...$ while the theoretical one, calculated from Eq. (\ref{relacion2}), is $\gamma_{q_{esc}} = 0.2223...$.}
\end{figure}

\begin{figure}[h]
\centering \includegraphics[width=8.5 cm,angle=0]{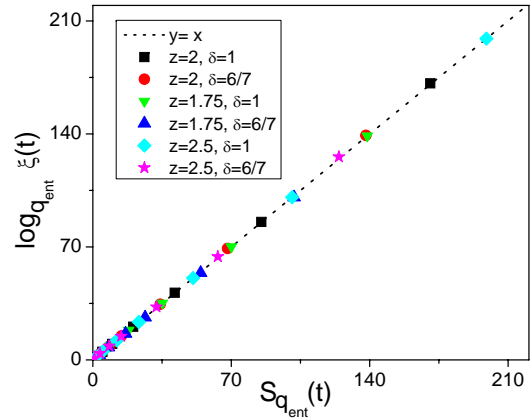}
\caption{Sensitivity to initial condition versus entropy production, see Eqs. (16) and (17), for different values of $z$. For $z=2$ and $\delta=0$: $K_{q_{ent}}=\lambda_{q_{sen}}=1.32...$, and $q_{ent}=q_{sen}=0.244...$; while for $z=2$ and $\delta=6/7$: $\gamma_{q_{esc}}=0.222...$, from Eq. (\ref{relacion2}), $K_{q_{ent}} = 1.1012...$ and ${q_{ent}}=0.0919... $, from Eqs. (\ref{pesinEscape}) and (\ref{q_pesinEscape2}). Similar results can be obtained for the other values of $z$. The holes are uniformly distributed in the line $y=0$. The continuous line correspond to a fit with a  slope 1.004..., numerically very close to unity, as expected. 
These examples neatly illustrate the validity of Eq. (17):  the ordinate corresponds to $(\lambda_{q_{sen}}-\gamma_{q_{esc}})~t$, and the abscissa corresponds to $K_{q_{ent}}~t$.}
\end{figure}

Then, if we introduce a fraction of $(1-\delta)$ traps (or holes) in the domain $-1 \le x \le +1$ , equidistantly  distributed in the line $y=0$, the system will suffer a decrease in the number of trajectories in a proportion given by $\delta^i$. The number of occupied bins, at time $t=2^i-1$, will be given by
\begin{eqnarray}
W(t)&=&W(2^i-1)=(1+t)^{\lambda_{q_{sen}}+\frac{\ln~\delta}{\ln~2}}.
\end{eqnarray}

Finally, after a straightforward calculation of the entropy production with

\begin{eqnarray}
q_{ent}&=&1-\frac{1}{\lambda_{q_{sen}}+\frac{\ln(\delta)}{\ln(2)}},
\end{eqnarray}
we obtain a generalization of the Pesin relation of rate parameters, including power-law escaping  Eq. (\ref{pesinEscape}) where we have used Eq. (\ref{q_pesinEscape2}), and
\begin{equation}
\gamma_{q_{esc}}=-\frac{\ln~\delta}{\ln~2} \label{relacion2}\,.
\end{equation}
In Fig. 1 we illustrate Eq. (\ref{exponentescape}), and in Fig. 2 we show a corroboration of equality (\ref{pesinEscape}). \\

\section{Final remarks}

To conclude, let us mention that one of the important aspects of dynamical systems with
escape is their connection with physical scenarios that present
boundaries in its phase space \cite{Gaspard98}. Our results show that, at the edge of chaos,
a thermodynamical-like quantity (entropy production) and a dynamical quantity (sensitivity to the initial conditions) are
connected even in the presence of escape. This type of connection, well known for chaotic
dynamical systems, is then valid, and natural, when considering a suitable entropy that takes into account the particular regime
(multifractal structure in phase space, power-law escaping, etc) under which the
system evolves.

The logistic map dynamics at the Feigenbaum point has
been recently studied in much detail \cite{BaldovinRobledo2004}. Its structured
behavior makes possible an analytic treatment. The results presented here
used this amenable characteristic of this particular map, but our main
point goes beyond it.
Our results would be applicable in principle to any system where both trajectory divergence within the
phase space, and escaping to outside of the system, follow power laws.
Typical examples would be deterministic anomalous diffusion in Lorenz gas, in
Hamiltonian systems (e.g., those numerically explored in \cite{SiopisKandrupContopoulosDvorak1996}), and spatio-temporal intermittency in dissipative systems.
As pointed out, our example, at the onset of chaos, can be an
archetypical illustration where the phenomenology discussed here occurs, but
also more complicated systems where attractors (see also \cite{Umarov}) with low-dimensional
behavior and its transients, that connect them, are embedded in a
larger system.

\textbf{Acknowledgements.} MAF thanks Prof. R. Speranza for fruitful discussions related to one-dimensional discrete dynamics. We acknowledge partial financial support by CNPq and Faperj (Brazilian agencies).

\end{document}